\documentstyle[multicol,psfig,prc,aps]{revtex}
\begin{document}
\draft
%
%
\def\mys#1{\Sigma_{ } { }_{\bf #1}}
\def\ro{\rho_{\rm o}}
\def\reff{$R_{eff}$} 
\def\lp{\left(} 
\def\rp{\right)}  
\def\bc{\begin{center}}
\def\ec{\end{center}}
\def\vp{\vspace*{0.1cm}}
\def\vm{\vspace*{-0.2cm}}
\def\CF{{\it CF}\,\,} 
\def\CFful{{Cooper-Frye}\,\,}
\def\CO{{\it CO}\,\,} 
\def\COful{{\it cut-off}\,\,} 
%
%
%
%

\title{
Relativistic Kinetic Equations for Finite Domains
and Freeze-out Problem
}

\author{\bf K.A. Bugaev }
\address{
Bogolyubov Institute for Theoretical Physics,
Kiev, Ukraine\\
Gesellschaft f\"ur Schwerionenforschung (GSI), Darmstadt, Germany
}

\date{\today}

\maketitle

\noindent
\begin{abstract}
The relativistic kinetic equations for the two 
domains separated by the hypersurface with both space- and time-like
parts are derived. The particle exchange between the domains separated by the time-like boundaries
generates source terms and modifies the collision term of the kinetic equation.
The correct hydrodynamic equations for the ``hydro+cascade'' models are obtained 
and their differences from existing freeze-out models of the hadronic matter  are discussed.     
\end{abstract}

\vspace*{0.2cm}

\noindent
\hspace*{1.9cm}\begin{minipage}[t]{14.cm}
\end{minipage}

\begin{multicols}{2}

\vspace*{-1.cm}

\section{Introduction}

\vm
\vm

In recent years an essential progress has been achieved
in our understanding of the freeze-out problem  in relativistic hydrodynamics,
i.e. how to convert hydrodynamic solution into free streaming particles.
Thus, in works \cite{BUG:96}
the correct generalization of the famous Cooper-Frey 
formula \cite{COOP:74} for the time-like hypersurfaces,
the \COful formula, was derived for the first time.
This was a necessary, but yet not sufficient step to formulate
the relativistic hydrodynamics without causal paradoxes.
The main problem was to formulate the energy-momentum and particle
conservation not only for the expanding fluid alone, but to extend it onto
the system consisting of the fluid and the gas of free streaming particles which
are emitted from the freeze-out hypersurface.
The principal solution of this task was given in \cite{BUG:96} and its
further analysis is presented 
in \cite{BUG:99b}. 
The advantage of this approach is the absence of the logical and causal paradoxes
usually arising, if emission of particles happens from the time-like hypersurfaces \cite{BUG:96,BUG:99b}. 
The disadvantage of this approach lies in its tremendous numerical complexity.  

Numerous attempts to improve  and 
to develop this approach further using primitive kinetic models \cite{MAGAS:99,MAGAS:99b}   
were not very successful so far.
Very recently a more fundamental treatment \cite{SIN:02} based on the analysis of the Boltzmann equation
was suggested. This approach, however, did not overcome  the usual difficulty of transport equations
in describing the phase transition phenomenon. 
This  difficulty has been overcome naturally within the ``hydro + cascade'' models 
suggested in Ref. \cite{BD:00} (BD) and further developed in \cite{SHUR:01} (TLS).
The latter models assume that  the nucleus-nucleus collisions 
proceed in three stages: hydrodynamic 
expansion (``hydro'') of quark gluon plasma (QGP), {\bf phase} transition from QGP to
hadron gas (HG) and the stage of hadronic
rescattering and resonance decays (``cascade''). The switch from hydro to
cascade modeling takes place 
at the boundary between the mixed  and hadronic phases. 
The spectrum of hadrons
leaving this hypersurface of the QGP--HG transition is taken as input for the
cascade.

Evidently, such an approach incorporates the most attractive features of both
hydrodynamics, which describes the QGP--HG phase transition very well,
and cascade, which works better during hadronic rescattering.
However, both the BD and TLS models face some 
principal  difficulties  
which cannot be ignored. Thus, within the BD approach 
the initial distribution for cascade is found by the \CFful formula \cite{COOP:74},
which takes into account particles with all possible velocities, 
whereas in the TLS model the initial cascade distribution is given by the \COful formula
\cite{BUG:96,BUG:99b},
which accounts for only those particles that can leave the phase boundary. 
As shown below  the \CFful formula will lead to  causal and
mathematical problems in the present version of BD model because the
QGP--HG phase boundary inevitably has time-like parts.
On the other hand the TLS model from the beginning does not conserve
energy, momentum and number of charges and this, as demonstrated later,
is due to the fact that the equations of motion used in \cite{SHUR:01} are not complete and, hence,
 should be modified.

The main  difficulty of the ``hydro + cascade'' approach looks very similar to 
the freeze-out problem in relativistic hydrodynamics.
In both cases the finite domains (subsystems) have time-like boundaries
through which the exchange of particles is occurring and this should be taken into account.
In relativistic hydrodynamics the problem was solved by the constraints which appeared on the freeze-out 
hypersurface and provided the global energy-momentum and charge conservation \cite{BUG:96,BUG:99b}.
Similarly, in kinetic theory one has to modify the transport  equations 
by the source terms which describe the
exchange of the particles on the time-like parts of the boundary between domains.
Therefore, we shall consider the two semi-infinite domains which are 
separated by the hypersurface $\Sigma^*$ of general type and rederive the 
kinetic equations for this case in Sect.~II. 
In Sect.~III the modification of the collision terms 
is found  and the relation between the system obtained and
the Boltzmann equation is discussed.
%
The correct equations of motion for the ``hydro + cascade'' approach
are analyzed in Sect.~IV.
%

\vm
\vm
 
\section{Drift Term for Semi-Finite Domain}
  
\vm
\vm

Let us consider the two semi-finite domains, ``in'' and ``out'',
separated by the hypersurface $\Sigma^*$ which in (3+1) dimensions will be parameterized
as $t = t^*(\bar{x}) = x_0^*(\bar{x})$. 
The distribution function $\phi_{in}(x,p)$  for $t \le t^*(\bar{x})$  
belongs to the ``in'' domain,
whereas $\phi_{out}(x,p)$ denotes the distribution function of the ``out'' domain 
for $t \ge t^*(\bar{x})$.
Throughout this work it is assumed that the initial conditions for 
$\phi_{in}(x,p)$ are given, whereas the initial  conditions for
$\phi_{out}(x,p)$ are not specified yet and will be the theme
of the subsequent discussion.
For simplicity we  consider a classical gas of point-like Boltzmann particles.

Similarly to Ref. \cite{GROOT} we  derive the kinetic equations
for $\phi_{in}(x,p)$ and $\phi_{out}(x,p)$ from the requirement of  
particle number conservation. 
Therefore, the particles leaving one domain (and crossing hypersurface $\Sigma^*$) 
should be subtracted from the corresponding distribution function and
added to the other one.
Now we consider the closed hypersurface  of the ``in'' domain, $\Delta x^3$,  which consists of
two semi-planes $\sigma_{t1}$ and $\sigma_{t2}$ of constant time $t1$ and $t2$, respectively,
that are connected from $t1$ to $t2 > t1$ by the part of the boundary $\Sigma^*(t1,t2)$.
The original number of particles on the hypersurface $\sigma_{t1}$ is given by 
the standard expression \cite{GROOT}

\vm
\begin{equation}\label{one}
N_1 = - \int\limits_{\sigma_{t1}} d \Sigma_\mu \frac{d^3 p}{p^0}~ p^\mu~ \phi_{in}(x,p)\,, 
\end{equation}
\vm
\vm
 
\noindent
where $d \Sigma_\mu$ is  the external normal vector to $\sigma_{t1}$ and, hence, 
the product $p^\mu d \Sigma_\mu \le 0$ is non-positive. 
It is clear that without collisions these particles can 
cross either hypersurface $\sigma_{t2}$ or $\Sigma^*(t1,t2)$.
The corresponding numbers of particles are as follows

\vm
\begin{eqnarray}\label{two}
N_2  \hspace*{0.0cm}=   \hspace*{-0.1cm}\int\limits_{\sigma_{t2}} \hspace*{-0.1cm}&& d \Sigma_\mu \frac{d^3 p}{p^0}~ p^\mu~ \phi_{in}(x,p)\,,\\ 
\label{three}
N_{loss}^*  = \hspace*{-0.2cm}\int\limits_{\Sigma^*(t1,t2)}\hspace*{-0.4cm}&&  d\Sigma_\mu 
\frac{d^3 p}{p^0}~ p^\mu~ \phi_{in}(x,p) ~\Theta(p^\nu d\Sigma_\nu) \,.
\end{eqnarray}
\vm
\vm

\noindent
The $\Theta$-function in the {\it loss} term (\ref{three}) is very important
because it accounts for the particles leaving the ``in'' domain  (see also
discussion in \cite{BUG:96,BUG:99b}). 
For the space-like parts of the hypersurface  $\Sigma^*(t1,t2)$ which are 
defined by negative  sign $ds^2 < 0$ of 
the element square, $ds^2 = dt^*(\bar{x})^2 - d\bar{x}^2 $, 
the product $p^\nu d\Sigma_\nu > 0$ is always positive and, therefore, 
particles with all possible  momenta can leave the ``in'' domain through
the $\Sigma^*(t1,t2)$.
For the time-like parts of $\Sigma^*(t1,t2)$ (with sign $ds^2 > 0$) 
the product $p^\nu d\Sigma_\nu $ can have either sign, and the $\Theta$-function
{\it cuts off} those particles which return to the ``in'' domain.

Similarly one has to consider the particles coming to the  ``in'' domain 
from outside. This is possible through the time-like parts of the hypersurface $\Sigma^*(t1,t2)$, 
if particle momentum satisfies  the inequality $ - p^\nu d\Sigma_\nu > 0$. 
In terms of the external normal $d \Sigma_\mu$ with respect to the ``in'' domain 
(the same as  in (\ref{three}))
the number of gained particles  
\begin{equation}\label{four}
N_{gain}^*  = - \int\limits_{\Sigma^*(t1,t2)}\hspace*{-0.4cm}  d\Sigma_\mu
\frac{d^3 p}{p^0}~ p^\mu~ \phi_{out}(x,p) ~\Theta(-p^\nu d\Sigma_\nu) \,
\end{equation}
\vm
\vm
 
\noindent
is, evidently, non-negative (compare it with contribution of  feed-back particles in \cite{BUG:96}).
Since the total number of particles is conserved, i.e. 
$N_2 = N_1 - N_{loss}^* + N_{gain}^*$, one can use the Gauss theorem
to rewrite the obtained integral over the closed hypersurface $\Delta x^3$ 
as the integral over $4$-volume $\Delta x^4 $  surrounded  
by $\Delta x^3$

\vm
\begin{eqnarray}
&&\int\limits_{\Delta x^4} \hspace*{-0.1cm}  d^4 x
\frac{d^3 p}{p^0}~ p^\mu  ~{\partial}_\mu ~ \phi_{in}(x,p) = \nonumber \\ 
\label{five} 
\vm
&&\int\limits_{\Sigma^*(t1,t2)}\hspace*{-0.4cm}  d\Sigma_\mu
\frac{d^3 p}{p^0}~ p^\mu \biggl[\phi_{in}(x,p) - \phi_{out}(x,p) \biggr] \Theta(-p^\nu d\Sigma_\nu) \,.
\end{eqnarray}
\vm
\vm
 
\noindent
Note that in contrast to the usual case \cite{GROOT}, i.e. in the absence of boundary $\Sigma^*$,   
the r.h.s of Eq. (\ref{five})  does not vanish identically.

The r.h.s of Eq. (\ref{five}) can be transformed further to the $4$-volume integral in the following
sequence of steps. First we express the integration element $d\Sigma_\mu$
via the normal vector $n^*_\mu$ as follows $(dx^j > 0,$ for $ j =1,2,3)$ 

\vm
\begin{equation}\label{six}
 d\Sigma_\mu = n^*_\mu dx^1 dx^2 dx^3; 
 \quad  n^*_\mu \equiv \delta_{\mu 0} - \frac{ \partial t^*(\bar{x}) }{\partial x^\mu} (1 - \delta_{\mu 0} )\,, 
\end{equation}
\vm
\vm
 
\noindent
where $\delta_{\mu \nu}$ denotes the Kronecker $\delta$-function. 
Then, using identity  $\int\limits_{t1}^{t2} dt\, \delta (t - t3) = 1$  
for the Dirac $\delta$-function
with
$t1 \le t3 \le t2$, we rewrite the r.h.s. integral in (\ref{five}) as 

\vm
\begin{equation}\label{seven} 
\int\limits_{\Sigma^*(t1,t2)}\hspace*{-0.4cm}  d\Sigma_\mu \cdots \equiv 
\int\limits_{V^4_\Sigma} d^4 x~\delta (t - t^*(\bar{x}) )~ n^*_\mu \cdots\,, 
\end{equation}
\vm
\vm
 
\noindent
where the $4$-dimensional volume $V^4_\Sigma$ is a direct product of the $3$- and $1$-dimensional
volumes $\Sigma^*(t1,t2)$ and $(t2-t1)$, respectively. 
Evidently, the Dirac $\delta$-function allows us to extend integration in (\ref{seven}) to the
unified $4$-volume  $V^4_U = \Delta x^4 \cup V^4_\Sigma$ of $\Delta x^4$ and $V^4_\Sigma$. 
Finally, with the help of notations

\begin{equation}\label{eight}
\Theta_> \equiv \Theta (t - t^*(\bar{x}) ); \quad \Theta_< \equiv 1- \Theta_> 
\end{equation}
it is possible to extend the l.h.s. integral in Eq. (\ref{five}) from $\Delta x^4$ to
$ V^4_U$.
Collecting all the above results, from Eq. (\ref{five}) one obtains

\vm
\begin{eqnarray}
&&\int\limits_{ V^4_U} \hspace*{-0.1cm}  d^4 x
\frac{d^3 p}{p^0}~ \Theta_<~ p^\mu  ~{\partial}_\mu ~ \phi_{in} = \nonumber \\
\label{nine}
\vm
&&\int\limits_{V^4_U}\hspace*{-0.1cm}  d^4 x
\frac{d^3 p}{p^0}~ p^\mu n^*_\mu \biggl[\phi_{in} - \phi_{out} \biggr] \Theta(-p^\nu n^*_\nu) 
~\delta (t - t^*(\bar{x}) ) \,.
\end{eqnarray}
\vm
\vm
 
\noindent
Since volumes $\Delta x^4$ and $V^4_U$ are arbitrary, one obtains 
the collisionless kinetic equation for the distribution function of the ``in'' domain 

\vm
\begin{eqnarray}
&& \Theta_<~ p^\mu  ~{\partial}_\mu ~ \phi_{in} (x,p) = \nonumber \\
\label{ten}
&&  p^\mu n^*_\mu \biggl[\phi_{in}(x,p) - \phi_{out}(x,p) \biggr] \Theta(-p^\nu n^*_\nu)
~\delta (t - t^*(\bar{x}) ) \,.
\end{eqnarray}
\vm
\vm
 
\noindent
Similarly one can obtain the equation for the distribution 
function of the ``out'' domain
\begin{eqnarray}
&& \Theta_>~ p^\mu  ~{\partial}_\mu ~ \phi_{out} (x,p) = \nonumber \\
\label{eleven}
&&  p^\mu n^*_\mu \biggl[\phi_{in}(x,p) - \phi_{out}(x,p) \biggr] \Theta(p^\nu n^*_\nu)
~\delta (t - t^*(\bar{x}) ) \,
\end{eqnarray}
with the same  
normal vector $n^*_\nu$  as  in Eq. (\ref{ten}).
Note the asymmetry between the r.h.s. of Eqs. (\ref{ten})
and (\ref{eleven}): for the space-like parts of hypersurface $\Sigma^*$ 
the  source term with $\Theta(-p^\nu n^*_\nu) $ vanishes identically because   $p^\nu n^*_\nu > 0$.
This reflects the  causal properties of the equations above:  
propagation of particles faster than light is forbidden, and hence no particle
can (re)enter the ``in'' domain. 
 

\vm
 
\section{ Collision Term for Semi-Finite Domain}
  
\vm

Since in the general case $\phi_{in}(x,p) \neq  \phi_{out}(x,p)$ on  $\Sigma^*$,  the r.h.s. of 
Eqs. (\ref{ten}) and (\ref{eleven}) cannot vanish simultaneously on this hypersurface. Therefore,
functions $\Theta_<^*  \equiv \Theta_<|_{\Sigma^*} \neq 0$ and 
$ \Theta_>^*  \equiv \Theta_>|_{\Sigma^*} \neq 0$ do not vanish simultaneously on  $\Sigma^*$ as well. 
For definiteness it is assumed that 
\vm
\begin{equation}\label{twelve}
\Theta_<^* = \Theta_>^* = \Theta (0) = \frac{1}{2}\,,
\end{equation}

\vm

\noindent
but the final results are independent of this choice.

Now the collision terms for Eqs. (\ref{ten}) and (\ref{eleven}) can be readily obtained.  
Adopting the usual assumptions on the distribution functions \cite{GROOT}, one can   
repeat the standard derivation of the collision terms \cite{GROOT} and get the desired expressions. 
We shall not recapitulate this standard part, but only discuss  how to modify the derivation for 
our purpose.
First of all, one has to start the derivation in the $\Delta x^4$ volume of the ``in'' domain and then
extend it onto the unified $4$-volume $V^4_U = \Delta x^4 \cup V^4_\Sigma$ similarly to the preceding section.
Then the first part of the collision term for Eq. (\ref{ten}) reads as 
\begin{eqnarray}\label{thirteen}
C_<^{I} (x,p) & = & \Theta_<^2 \left( I_G [\phi_{in}, \phi_{in}] - I_L [\phi_{in}, \phi_{in}] \right) 
\,, \\
\label{fourteen}
I_G [\phi_{A}, \phi_{B}] & \equiv & \frac{1}{2} \int D^9 P~  
\phi_{A}(p^{\prime}_1 )~ \phi_{B}(p_1^{\prime})~ W_{p,p_1^{} | p^{\prime}p_1^{\prime}} 
\,, \\ 
\label{fifteen} 
I_L [\phi_{A}, \phi_{B}] & \equiv & \frac{1}{2} \int D^9 P~
\phi_{A}(p)~ \phi_{B}(p_1)~ W_{p,p_1^{} | p^{\prime}p^{\prime}_1}\,,
\end{eqnarray}
\vm
\vm

\noindent
where the invariant measure of integration is denoted as
$ D^9 P \equiv \frac{d^3 p_1}{p^0_1} \frac{d^3 p^{\prime} }{p^{\prime 0}} 
\frac{d^3 p^{\prime}_1 }{p^{\prime 0}_1} $ and $W_{p,p_1^{} | p^{\prime}p^{\prime}_1}$ 
is the transition rate in the elementary  reaction   
with energy-momentum conservation given in the form
$p^\mu + p_1^\mu = p^{\prime \mu} + p^{\prime \mu}_1$.
The r.h.s. of  (\ref{thirteen}) contains the square of $\Theta_<$-function 
because the additional $\Theta_<$ accounts for the fact that 
on the boundary hypersurface $\Sigma^*$ one has to take  only one half 
of the traditional collision term (due to Eq. (\ref{twelve}) only one half of 
$\Sigma^*$ belongs to the ``in'' domain). 
It is easy to understand that
on $\Sigma^*$ 
the second part of the collision term 
(according to Eq. (\ref{twelve})) is defined by the 
collisions between particles of ``in'' and ``out'' domains
\begin{equation}\label{sixteen}
C_<^{II} (x,p)  =  \Theta_< \Theta_>  \left( I_G [\phi_{in}, \phi_{out}] - I_L [\phi_{in}, \phi_{out}] \right)
\,. 
\end{equation}
Combining results (\ref{ten}), (\ref{thirteen}) and (\ref{sixteen}), we obtain the kinetic
equation for the semi-finite ``in'' domain 
\begin{eqnarray}
&& \Theta_<~ p^\mu  ~{\partial}_\mu ~ \phi_{in} (x,p) =  C_<^{I} (x,p) + C_<^{II} (x,p) + \nonumber \\
\label{seventeen}
&&  p^\mu n^*_\mu \biggl[\phi_{in}(x,p) - \phi_{out}(x,p) \biggr] \Theta(-p^\nu n^*_\nu)
~\delta (t - t^*(\bar{x}) ) \,.
\end{eqnarray}
\vm
\vm
 
\noindent
The corresponding equation for the ``out'' domain
\begin{eqnarray}
&& \Theta_>~ p^\mu  ~{\partial}_\mu ~ \phi_{out} (x,p) =  C_>^{I} (x,p) + C_>^{II} (x,p) +  \nonumber \\
\label{eighteen}
&&  p^\mu n^*_\mu \biggl[\phi_{in}(x,p) - \phi_{out}(x,p) \biggr] \Theta(p^\nu n^*_\nu)
~\delta (t - t^*(\bar{x}) ) \,
\end{eqnarray}
\vm
\vm

\noindent
can be derived similarly. In (\ref{eighteen}) we used the evident 
notations  
$C_>^{I} \equiv \Theta_>^2 \left( I_G [\phi_{out}, \phi_{out}] - I_L [\phi_{out}, \phi_{out}] \right) $
and
$C_>^{II} \equiv \Theta_< \Theta_> \left( I_G [\phi_{out}, \phi_{in}] - I_L [\phi_{out}, \phi_{in}] \right) $.

For the continuous distribution functions on $\Sigma^*$, i.e. $\phi_{out}|_{\Sigma^*} = \phi_{in}|_{\Sigma^*}$,   
the source terms in r.h.s. of Eqs. (\ref{seventeen}) and (\ref{eighteen}) 
vanish and one
recovers the Boltzmann equations.
With the  help of the evident relations

\vm
\begin{eqnarray}\label{nineteen}
&&- {\partial}_\mu ~ \Theta_< = {\partial}_\mu ~ \Theta_> = n_\mu^*~ \delta (t - t^*(\bar{x}) )\,, \\
\label{tw}
&&C_<^{I} + C_<^{II} + C_>^{I} + C_>^{II} =  
I_G [\Phi, \Phi] - I_L [\Phi, \Phi]\,, 
\end{eqnarray} 
\vm
\vm
 
\noindent
where the notation $\Phi(x,p) \equiv \Theta_<~\phi_{in}(x,p) + \Theta_>~\phi_{out}(x,p) $ 
is used, one can get the following result for the sum of Eqs. (\ref{seventeen}) and (\ref{eighteen}) 
\begin{equation}\label{twone}
p^\mu  ~{\partial}_\mu ~ \Phi (x,p) = I_G [\Phi, \Phi] - I_L [\Phi, \Phi]\,.
\end{equation}
\vm
\vm
\vm
 
\noindent
In other words, the usual Boltzmann equation follows from
the system of Eqs. (\ref{seventeen}) and (\ref{eighteen})
automatically without {\it any assumption} about the behavior 
of $\phi_{in}$ and $\phi_{out}$ on the boundary hypersurface
$\Sigma^*$.
In fact the system (\ref{seventeen}, \ref{eighteen}) generalizes 
the relativistic kinetic equation to the case of the strong 
temporal and spatial inhomogeneity, i.e.,
if $\phi_{in}(x,p) \neq \phi_{out}(x,p)$ on $\Sigma^*$.
Of course, one has to be extremely careful while discussing
the strong temporal inhomogeneity (or discontinuity on the space-like parts of $\Sigma^*$) 
such as the so called {\it time-like shocks} \cite{TIMESHOCK}
because their existence may  
contradict to the usual assumptions 
adopted for distribution function. 
Therefore, in what follows we shall discuss exclusively
the spatial inhomogeneities or discontinuities on the time-like parts of $\Sigma^*$ which are less restrictive
because in some sense the equations above are delocalized in  space.

From the system (\ref{seventeen}), (\ref{eighteen}) it is possible to derive the 
macroscopic equations of motion by multiplying the corresponding equation with $p^\nu$ 
and integrating it over the invariant measure. Thus  Eq. (\ref{seventeen})
generates the following expression ($T^{\mu \nu}_{A} \equiv \int \frac{d^3 p }{p^ 0}~ p^\mu p^\nu \phi_{A}(x,p) $)

\vm
\begin{eqnarray}
&& \Theta_<~ {\partial}_\mu ~ T^{\mu \nu}_{in} =  \int \frac{d^3 p }{p^ 0}~  p^\nu  
C_<^{II} (x,p) + \nonumber \\
\label{twtwo}
&& \int \frac{d^3 p }{p^ 0}~ p^\nu  p^\mu n^*_\mu 
\biggl[\phi_{in} - \phi_{out} \biggr] \Theta(-p^\rho n^*_\rho)
~\delta (t - t^*(\bar{x}) ) \,.
\end{eqnarray}
\vm
\vm
 

\noindent
Similarly to the usual Boltzmann equation
the momentum integral of the collision term $C_<^{I}$ vanishes due to its symmetries,
but it can be shown 
that the integral of the second collision term $C_<^{II}$ does   
not vanish because it involves two different (and not identical) 
distribution functions. 
The corresponding equation for the ``out'' domain follows similarly

\vm
\begin{eqnarray}
&& \Theta_>~ {\partial}_\mu ~ T^{\mu \nu}_{out} =  \int \frac{d^3 p }{p^ 0}~  p^\nu
C_>^{II} (x,p) + \nonumber \\
\label{twthree}
&& \int \frac{d^3 p }{p^ 0}~ p^\nu  p^\mu n^*_\mu
\biggl[\phi_{in} - \phi_{out} \biggr] \Theta(p^\rho n^*_\rho)
~\delta (t - t^*(\bar{x}) ) \,.
\end{eqnarray}
\vm
\vm
 
  
\noindent
Note that similar equations (with $\delta$-like term)  
first were obtained within the relativistic hydrodynamics in  \cite{BUG:96}. 

\vm
\vm
 
\section{Discussion}
  
\vm
\vm

It is clear that  Eqs. (\ref{seventeen}), (\ref{eighteen}), (\ref{twtwo}) and (\ref{twthree})
remain valid for the finite domains as well.
With their help 
we are ready to analyze the ``hydro+cascade'' models. 
In the TLS model the \COful formula relates $\phi_{in}$ ($\equiv$ hydro, Eq. (\ref{twtwo})) and 
$\phi_{out}$ ($\equiv$ cascade, Eq. (\ref{eighteen})) on $\Sigma^*$
as follows 
\begin{equation}\label{twfoure}
{\rm TLS:} \quad  \phi_{out}\biggl|_{\Sigma^*} \hspace*{-0.2cm}  = \Theta(p^\rho n^*_\rho)~ \phi_{out}\biggl|_{\Sigma^*} 
\hspace*{-0.2cm} =
\Theta(p^\rho n^*_\rho)~ \phi_{in}\biggl|_{\Sigma^*} \hspace*{-0.1cm}\,,
\end{equation}
\vm
 
\noindent
i.e., for the space-like parts of hypersurface $\Sigma^*$ these functions are identical,
whereas for the time-like parts of $\Sigma^*$ there are no returning particles to the ``in'' domain.
In this case the source term  in cascade Eq. (\ref{eighteen}) is zero, while 
the source term in hydro Eq. (\ref{twtwo}) does not vanish on the time-like parts of the boundary
$\Sigma^*$.
Therefore, the main defect of the TLS model is not even the energy-momentum non-conservation, but the
incorrect hydrodynamic equations. The absence of the $\delta$-like 
source term in \cite{SHUR:01} breaks the conservation laws (evidently,
the system (\ref{twtwo},\ref{twthree}) obeys the conservation laws), but its inclusion  
into consideration will inevitably change the hydrodynamic solution of Ref. \cite{SHUR:01}. 
The full analysis of the possible solutions of the systems 
(\ref{seventeen},\ref{eighteen}) and (\ref{twtwo},\ref{eighteen}) requires 
a special consideration. We only mention that from the negative sign of the 
TLS source term in the r.h.s. of (\ref{twtwo}) for equal indices $\nu = \mu$ one immediately can deduce that 
such a correction to the hydro equations should increase the degree of the
fluid rarefaction in comparison with the standard hydrodynamic expansion. 
It is, therefore, quite possible that such a source term will generate
a discontinuity between ``in'' and ``out'' domains. 
In the thermodynamically normal media \cite{BUG:88} the rarefaction
shocks are mechanically unstable. However, 
it is well know that on the phase transition boundary between QGP and HG
the properties of the mixed phase are thermodynamically anomalous \cite{BUG:88}
and the usual rarefaction shocks are possible.  
Another possibility is the occurrence of 
the new type of the discontinuity, the {\it freeze-out shock}  suggested
in Refs. \cite{BUG:96,BUG:99b}, where the post freeze-out state is
described by the \COful distribution and, hence, is very similar
to the TLS ansatz.
It is clear that in both cases the additional rarefaction will reduce
the mean transverse size and the life-time of the hadronizing QGP.

Let us consider briefly the BD approach.
Since in the BD model the hydro and cascade distributions  on 
$\Sigma^*$ are equal $\phi_{out}|_{\Sigma^*} = \phi_{in}|_{\Sigma^*}$,
the corresponding  source terms vanish in all equations. 
Therefore, at first glance the BD approach correctly 
conjugates  the hydro and cascade solutions
on the arbitrary hypersurface. 
For the oversimplified kinetics considered above it  is so.
However, the real situation differs essentially from our consideration. 
Thus, the hydro part in both the BD and TLS models is assumed to be in the local thermodynamic equilibrium,
whereas the matter in the cascade domain can be far from equilibrium \cite{BD:00,SHUR:01}
(this was, actually, the main reason why both groups decided to use the cascade). 
Consequently,  the BD transport equations for all hadrons are homogeneous in  the hydro domain,
whereas for the cascade domain they are inhomogeneous.
Since the initial BD cascade distribution on the time-like parts of $\Sigma^*$ contains 
the particles returning to the fluid domain  ($ p^\nu n^*_\nu < 0$), then these particles will move towards the
space-like parts of the hypersurface $\Sigma^*$ which are located 
inside of the light cone originated at each point of the time-like part of $\Sigma^*$. 
Then for each hadron  the inhomogeneous BD cascade equation will generate   
a different distribution function   
%
%
than the one already  obtained from the hydro equations 
(or homogeneous transport ones) on these space-like parts of $\Sigma^*$. 
Thus, one arrives at a causal paradox, the {\it recoil problem} \cite{BUG:96},  and at 
a mathematical inconsistency.

Evidently, the inclusion of the viscosity into the hydro equations (apart from its tremendous complexity
for the mixture of about hundred hadrons)
will not solve the problem  
because for the small deviations from equilibrium (and, hence, a small viscosity effect) the influence of the returning  
particles may remain essential. If, on the contrary, the matter 
in the hydro domain is far from equilibrium,
then the usage of the hydro equations becomes problematic.
Therefore, a more realistic way 
is 
to 
find the boundary conditions for $\phi_{in}$ and $\phi_{out}$ on the separating hypersurface $\Sigma^*$
form the system of kinetic equations  (\ref{seventeen},\ref{eighteen}),
and then to apply these boundary conditions to the system (\ref{seventeen},\ref{twtwo}) 
%
which ensures
the correct treatment of the relativistic nuclear collisions 
within the frame of the ``hydro+cascade'' model.


\noindent
{\bf  Acknowledgments.}  
The author 
 thanks 
 A. L. Blokhin, P. Braun-Munzinger, A. Dumitru, P. T. Reuter and D. H. Rischke 
 for very useful discussions and valuable comments. 

\def\np#1{{\it Nucl. Phys.} {\bf #1}}
\def\prl#1{{\it Phys. Rev. Lett.} {\bf #1}}
\def\jp#1{{\it J. of Phys.} {\bf #1}}
\def\zp#1{{\it Z. Phys.} {\bf #1}}
\def\pl#1{{\it Phys. Lett.} {\bf #1}}
\def\pr#1{{\it Phys. Rev.} {\bf #1}}
\def\hip#1{{\it Heavy Ion Physics} {\bf #1}}
\def\prep#1{{\it Phys. Rep.} {\bf #1}}
\def\preprint#1{{\it Preprint} {\bf #1}}

\end{multicols}
\end{document}